\RequirePackage{lineno}
\documentclass[twocolumn,showpacs,aps,prd,floatfix,altaffilletter,altaffillsymbol]{revtex4}
\usepackage{graphicx}
\usepackage{dcolumn}
\usepackage{amsmath}
\usepackage{epsfig}
\usepackage{subfigure}
\usepackage{longtable}
\usepackage{bm}
\usepackage{relsize}

\clearpage
\begin{document}
\title{  Studies of the X(3872) at Belle II}
\vspace{-6pt}
\author{  Elisabetta Prencipe    }
\address{   Justus-Liebig-University of Giessen, DE  }

\vspace{-12pt}

\begin{abstract}
  The X(3872) is one of the most puzzling resonances ever observed. First seen by the Belle Collaboration in 2003, it solicited the effort of a hundred of experimental physicists and dozens of theorists, who nowadays are trying yet to shed light on the nature of this peculiar resonant state. It was seen in several decay modes and different production mechanisms, and confirmed by several experiments, so it is well established, and recently  is addressed as the $\chi_{c1} (3872)$. Here we report about a re-discovery of the X(3872) with early Belle II data, and discuss plans for future measurements once the full integrated planned luminosity will be achieved by Belle II.
  
\end{abstract}
\maketitle

\section{Introduction}

The so-called X(3872) is an exotic resonant state that does not fit into potential models~\cite{models}. It was observed for the first time by the Belle Collaboration in the $B^\pm \rightarrow J/\psi \pi^+ \pi^- K^\pm$ decay channels~\cite{belle1}, by analyzing the $ J/\psi \pi^+ \pi^- $ invariant mass. This is one of the most cited articles ever published by Belle,  updated later in 2011~\cite{belle2}. Several experiments published on that~\cite{one, two, three, four, five, six, seven, eight, nine, ten, eleven}, also in different production mechanisms and other decay modes~\cite{x1, x2, x3, x4, x5, x6}. Nowadays the X(3872) is well established, and its interptetation is still puzzling because its quantum numbers do not fit into the potential models.

\section{State of the art}

The LHCb experiment definitively established that the X(3872) has $J^{PC}$ = 1$^{++}$~\cite{x7}, excluding in this way some hypotheses about its interpretation. The X(3872) can be unluckily interpreted as a standard charmonium state, due to its narrow width and strong isospin violation. The most suitable and preferred interpretations are nowadays charm molecule or tetraquark, but yet other hypotheses cannot be ruled out. In fact, among the possible explanations are those interpreting the X(3872) as a  hybrid state where the gluon field contributes to its quantum numbers, or a glueball without any valence quarks at all. A mixture of these explanations is also possible.

  The measurement of the X(3872) width could actually constrain theoretical models. The best value that Belle could measure as an upper limit (UL) at 90\% confidence level (c.l.) was 1.2 MeV~\cite{belle2}. No conventional hadron is expected to have such a narrow width in the charmonium spectrum. However, recently LHCb pushed further the investigation of the X(3872) width, in B decays~\cite{four} or inclusively~\cite{five}, always in the $J/\psi \pi^+ \pi^-$ final state, but using different data sets. By performing an analysis of the X(3872), in the assumption of a Breit-Wigner (BW) parameterization of its lineshape, LHCb established that the X(3872) width is equal to 1.39 MeV~\cite{five}. The reason of performing a simple BW fit is that it  neglects potential distortions.  A precise measurement of the X(3872) lineshape could help elucidate its nature.  LHCb then used also a Flatt\'e model, and the extremely challenging width value of 220 keV was measured~\cite{lhcb2}. The  LHCb results favour the interpretation of this state as a quasi-bound $D^0 \bar D^{*0}$ molecule. Further studies are ongoing.

  Both LHCb and Belle analyzed the invariant mass system of $J/\psi \pi^+ \pi^-$ in B decays, for the width measurement. The conclusion reported by the LHCb analyses~\cite{eleven, lhcb2} is that at the actual status of the art of this search there is no way to distinguish the Flatt\'e from the BW model. 

  The logic question could be whether exists or not  a  decay channel that could be more sensitive to the X(3872) width measurement, and if an experiment exists, which  can distinguish between different lineshape parameterizations. In other words, understanding the lineshape of the X(3872) plays a fundamental role in disclosing its nature.

  A leading role in undersanding the nature of the X(3872) is played by the analysis of the $X(3872) \rightarrow D^0 \bar D^{0*}$, which was started at Belle, but only 50 events were fitted over 657 fb$^{-1}$ data~\cite{x1}.  

  The analysis of the X(3872) in prompt production at FNAL and LHC showed interesting results:
  \begin{itemize}
   \item  production rate at Tevatron is too large by orders of magnitude for a X(3872) to be a weekly-bound charm molecule~\cite{tevatron, braaten}.
   \item re-scattering effects  could introduce additional interactions between D mesons in the final state, therefore the X(3872) production rate could enhance.
   \item re-scattering could be significant if the relative momenta of the D mesons are small, and at large transverse momenta. Therefore, measuring the $p_T$-dependence of the X(3872) production rate could give insights about the validity of the {\it charm-meson molecule} hypothesis.
     \item CMS has observed copious X(3872) produced in prompt processes rather than B mesons (only 26\% in B decays)~\cite{cms}: the predicted $p_T$-dependence of the X(3872) is actually larger than the measured rate, but fairly modeled. In addition, recent observation of the X(3872) in $B_s \rightarrow X(3872) \phi$ decays at CMS suggests another laboratory for studying its properties~\cite{cms2}.
    \end{itemize}

  LHCb recently scrutinized the nature of the X(3872) by studying its multiplicity dependent relative suppression compared to a conventional charmonium state, $i.e.$ $\psi(2S)$. In the hypothesis of the X(3872) being  a hadronic molecule, its radius should be large at the order of 10 fm, while in the hypothesis of a compact tetraquark it is supposed to be 1 fm~\cite{theo2}.   If one consider the decay of the X(3872) to $D^0 \bar D^{*0}$ mesons, the difference between the X(3872) and its decay products is found to be equal to 0.1 MeV/$c^2$.  LHCb has found that the X(3872) prompt ratio decreases with the multiplicity~\cite{lhcb3}, which means a stronger suppression of X(3872) over $\psi(2S)$ is observed. This argument is used against the charm molecule interpetation~\cite{theo3}.

  The Belle experiment has also given a remarkable contribution in trying to understand the properties of the X(3872), to better constrain theoretical models. In fact, it was measured:
  \begin{itemize}
  \item $\Delta M$, defined as the X(3872)  mass difference in B charged and B neutral decays. It is evaluated to be (-0.69 $\pm$ 0.97 $\pm$ 0.19) MeV/c$^2$, which is compatible with zero. This is against the quark-antiquark model.  
  \item R(X), defined as the ratio of the branching ratio of the charged and neutral B meson decays, where the X(3872) was observed. It was measured to be (0.50 $\pm$ 0.14 $\pm$ 0.04). In the molecular model, it should range in [0.06,0.29].   \item search for charged partners, which gave no positive outcome in the decays $B^0 \rightarrow K^- \pi^+ \pi^0 J/\psi$ and $B^+ \rightarrow K^0 \pi^+ \pi^0 J/\psi$.
   \item search for $B^{0,+} \rightarrow  D^0 \bar D^0 \pi^0 K^{0,+}$. The branching ratio of these 2 decay modes is found identical, within statistical error, then R(X) here is compatible with  1. Evidence for the $X(3872) \rightarrow D^0 \bar D^{0*}$ has been found at Belle.   
  \end{itemize}
  
  The analysis of the $X(3872) \rightarrow D^0 D^0 \pi^0$ is extremely interesting, since it shows sensitivity to the X(3872) width measurement. In fact, the difference between the X(3872) mass and that of its decay products in this case would be 7.05 MeV/$c^2$ ($D^0 \bar D^0 \pi^0$) and 0.1 MeV/$c^2$ ($D^0 \bar D^{0*}$). In order to perform this analysis, and experiment with  good photon reconstruction is required.

  \section{The Belle II experiment}

  The Belle II experiment is an asymmetric $e^+e^-$ collider, collecting data mostly at the center of mass energy of the $\Upsilon(4S)$, which decays to $B \bar B$ pairs. Spectroscopy analysis through B decays, or in the continuum, or via initial state radiation (ISR) are possible at B factories. So far Belle II collected 239 fb$^{-1}$ data in roughly one year of data taking, which corresponds to the integrated luminosity that the old Belle experiment collected in 4 years. The Belle II experiment can be considered as a major upgrade of the Belle experiment, and it is located at the same site, at KEKB (Tsukuba, Japan). The Belle II experiment is designed to reach an integrated luminosity of 50 ab$^{-1}$, for which both, the detector and the KEKB facility had to be upgraded.
  
\begin{figure}[htp!]
  \caption{ Integrated luminosity as function of the years, as planned in Belle II. The peak luminosity is shown under 2 different hypothesis: before and after the IR (interaction region) upgrade.}
  \includegraphics[width=7.5cm]{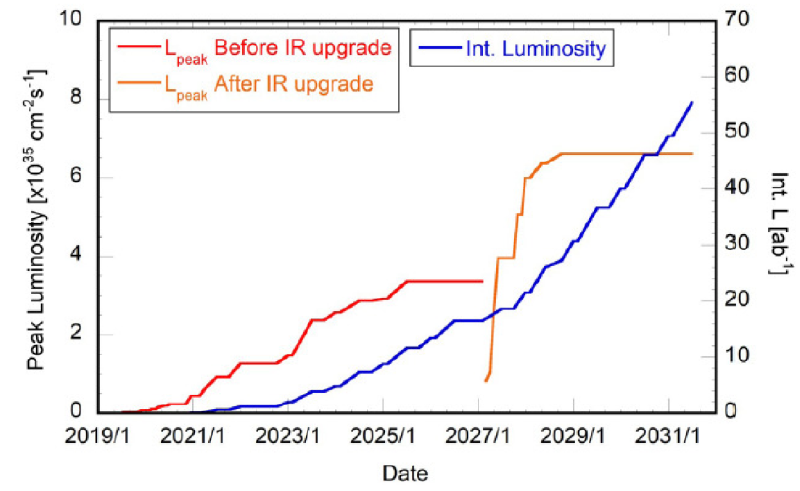}
  \label{unoeli}
\end{figure}

A plan for the future integrated lumnosity at Belle II is given in the scheme of Fig.~\ref{unoeli}.
  
  \section{Charmonium spectroscopy at Belle II}
  
Spectroscopy analysis through B decays, or in the continuum, or via initial state radiation (ISR) are possible at Belle II. The analysis of the X(3872) is a hot topic analysis of the Belle II charmonium spectroscopy program.
    
\begin{figure}[htp!]
  \caption{ MC simulations performed at Belle II: perspectives of the X(3872) width measurements, depending on the available integrated luminosity. The blue dots show that with a statistics of 50 ab$^{-1}$ data the limit that one could reach is 0.65 MeV
(5$\sigma$ effect), or 0.3 MeV with a 3$\sigma$ effect (red dots), or 189
keV as a new UL at 90\% c.l. (black dots). The old Belle UL corresponding to 1.2 MeV is shown as a bold horizontal red line.}
  \includegraphics[width=7.5cm]{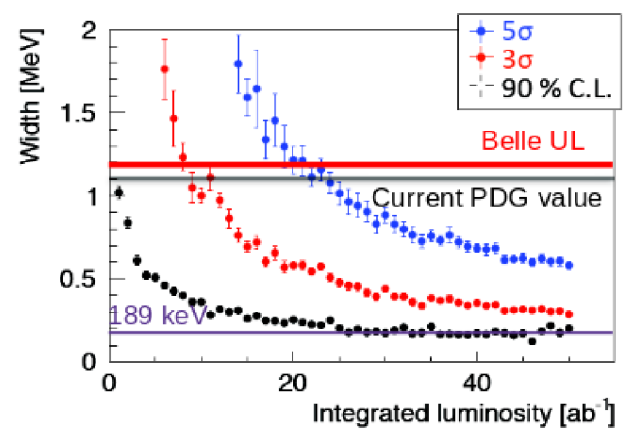}
  \label{dueeli}
\end{figure}

Our MC simulations demonstrated that the lower limit in the width measurement of the X(3872), when analyzing the $B \rightarrow D^0 \bar D^{0*} K$ decay channel, is 189 keV (see Fig.~\ref{dueeli}).

Further studies are ongoing, considering different models for the X(3827) lineshape, to understand if Belle II will be able in early future to discriminate between, $e.g$ the Flatt\'e and the BW parameterization, which so far LHCb is also not able to discriminate. Indeed LHCb in the hypothesis of Flatt\'e parameterization, published the impressive value of 220 keV for the X(3872) width~\cite{lhcb2}.
  The decay channel that will be under investigation in Belle II for the purpose of the measurement of the X(3872) width is $X(3872) \rightarrow D^0 \bar D^0 \pi^0$. The reason is that to constrain $D^{0*} \rightarrow D^0 \pi^0$ is a strong assumption, being unknown the pole position of the X(3872). In this case one makes the assumption that the X(3872) pole is above the $D^0 \bar D^{*0}$ threshold, for which we have got no confirmation so far. This assumption would exclude a priori a possible solution. By analyzing  $X(3872) \rightarrow D^0 \bar D^0 \pi^0$ all possibilities remain open.

  \section{Analysis of the $X(3872) \rightarrow J/\psi \pi^+ \pi^-$ at Belle II}
  
  With 62.8 fb$^{-1}$ re-processed Belle II data it was possibile to study $X(3872) \rightarrow J/\psi  \pi^+   \pi^-$ in B decays, and confirm the former Belle result. The analysis was conducted by analyzing the $B^{+,0} \rightarrow J/\psi \pi^+ \pi^- K^{+,0}$  channels.  As control sample, the analysis of the $\psi(2S) \rightarrow J/\psi \pi^+ \pi^-$ was performed. Particle identification was applied to leptons and pions involved in the decay channel under exam, and a standard mass window selection around the $J/\psi$ and $K^0_s$ masses is applied. $J/\psi$ is reconstructed to leptons ($e, \mu$), then mass constrained.

\begin{figure}[htp!]
    \caption{ Fit to Belle II data of the $J/\psi \pi^+ \pi^-$ invariant mass distribution in [left] $ B^\pm \rightarrow J/\psi \pi^+ \pi^- K^\pm$ and [right] $B^0 \rightarrow J/\psi \pi^+ \pi^- K^0_S$, using  62.8 fb$^{-1}$ data sample.}
    \includegraphics[width=8.5cm]{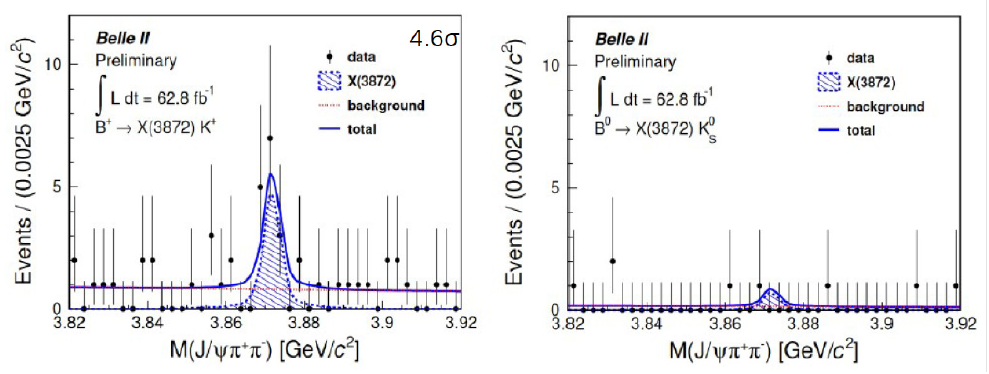}
    \label{hiratafig}
  \end{figure}

  Useful kinematic variables to study are $M_{bc}$ (beam-constrained mass) and $\Delta E$ (energy difference), defined as $M_{bc} = \sqrt(E^2_{beam}/c^4 - \lvert p_B/c \rvert ^2)$ and $\Delta E =  E_{beam} - E_B$, respectively. The continum suppression is guaranted by the condition $R_2 < 0.4$, where $R_2$ represents the Fox-Wolfram momentum of the second order, normalized to the zero order. The result of the unbinned maximum likelihood fit is reported in Fig.~\ref{hiratafig}. The study of the control sample reveals good agreement with the PDG value.

  With the statistics available for this study almost an observtion of the X(3872) is provided (4.6 $\sigma$ significance). The signal is efficiently reconstructed; 19.1\% reconstruction efficiency is quoted on the charged B channel. This preliminary analysis on  early Belle II data reveals an excellent agreement with
the old Belle analysis~\cite{belle2}, with improvement in term of reconstruction efficiency, and consequently fitted events.

\section{Conclusion}
The Belle II experiment is performing good, and so  far collected 239 fb$^{-1}$ data. Preliminary results on 62.8  fb$^{-1}$  data show the first re-discovery of the X(3872). We  are looking forward to collect the whole data set at the  c.m. energy of the $\Upsilon (4S)$, and repeat this interesting analysis in all possible decay modes at Belle II. A plan  to combine Belle and Belle II data for the investigation of the $X(3872) \rightarrow D^0 \bar D^0 \pi^0$ has been already approved.


\end{document}